\documentstyle[aps,prl,twocolumn,floats,epsfig]{revtex}
\begin{document}
\preprint{}
\title{Conjugation-Length Dependence of Spin-Dependent Exciton Formation Rates in $\Pi$-Conjugated Oligomers and Polymers}

\author{M. Wohlgenannt, X. M. Jiang and Z. V. Vardeny}
\address{Department of Physics, University of Utah,
115 South 1400 East, Salt Lake City, Utah 84112-0830}

\author{R. A. J. Janssen}
\address{Laboratory for Macromolecular and Organic Chemistry,
Eindhoven University of Technology, \\ PO Box 513, 5600 MB Eindhoven, The Netherlands}

\maketitle

\begin{abstract} 
We have measured the ratio, r = $\sigma_S/\sigma_T$ of the formation cross section, $\sigma$ of singlet ($\sigma_S$) and triplet ($\sigma_T$) excitons from oppositely charged polarons in a large variety of $\pi$-conjugated oligomer and polymer films, using the photoinduced absorption and optically detected magnetic resonance spectroscopies. The ratio r is directly related to the singlet exciton yield, which in turn determines the maximum electroluminescence quantum efficiency in organic light emitting diodes (OLED). We discovered that r increases with the conjugation length, CL; in fact a universal dependence exists in which $r^{-1}$ depends linearly on $CL^{-1}$, irrespective of the chain backbone structure. These results indicate that $\pi$-conjugated polymers have a clear advantage over small molecules in OLED applications. 
\end{abstract}

The efficiency of fluorescence-based organic light emitting diodes (OLED) is determined by the fraction of injected electrons (e) and holes (h) that recombine to form emissive spin-singlet excitons, rather than the non-emissive spin-triplet excitons. If the process by which these excitons form were spin-independent, then the maximum quantum efficiency, $\eta_{max}$ of OLEDs would be limited to 25\%\cite{review}, which is the statistical limit. The reason for the 25\% statistical limit is that the combination of two spin-1/2 particles gives four possible total spin states; three of which have total spin 1, only one is a singlet state. But recent reports have indicated that $\eta_{max}$ in $\pi$-conjugated OLEDs ranges between 22\% to 63\%\cite{rAlQ3,rmeasurement1,rmeasurement2,nature,Friendnature}, and the reason for this variation is under investigation. In particular the dependence of $\eta_{max}$ on the conjugation length (CL) was recently tested by measuring $\eta_{max}$ in a monomer and related polymer, and the possibility that $\eta_{max}$ increases with the CL in $\pi$-conjugated materials was advanced\cite{Friendnature}. 

For systems which are light emitting the quantum efficiency, $\eta_{EL}$ for electro-luminescence (EL), is $\eta_{EL} = \eta_1\eta_2\eta_3$, where $\eta_1$ is the singlet emission quantum efficiency, $\eta_2$ is the fraction of the total number of excitons that are singlets, and $\eta_3$ is the probability that the injected e and h find each other to form e-h pairs\cite{Bredas}. Since both $\eta_1$ and $\eta_3 <1$, it follows that $\eta_{EL} < \eta_2 = \eta_{max}$. We have recently developed\cite{nature} a spectroscopic technique based on photoinduced absorption (PA) and PA-detected magnetic resonance (PADMR) spectroscopies, which allows direct measurement of the ratio r = $\sigma_S/\sigma_T$ of the formation cross-section, $\sigma$ of singlet ($\sigma_S$) and triplet ($\sigma_T$) excitons from oppositely charged polarons in films of $\pi$-conjugated materials. In the limiting case that the spin relaxation rates are much faster than the exciton formation rates from e-h polaron pairs we showed that $\eta_{max} = (1+3r^{-1})^{-1}$\cite{nature}. Thus the study of the ratio r in organic materials also provides information about $\eta_{max}$ in OLEDs. Using this spectroscopic technique we could infer r\cite{nature}, and consequently $\eta_{max}$ in a variety of $\pi$-conjugated materials without the need to fabricate OLEDs based on the particular material as the active layer. We found that r depends on the material optical gap in a systematic but non-monotonic way, and the reason for this was investigated\cite{nature}. 

In the present work we measured the dependence of r, and consequently $\eta_{max}$ on the CL in a large variety of $\pi$-conjugated oligomer and polymer films using the spectroscopic method developed previously\cite{nature}. We indeed found that r increases with the CL; (here we denote with the number n a CL corresponding to the length of an oligothiophene with n "rings"), where $r \approx 1$ in small molecules, which are considered to have CL of $n \approx 1$. Moreover we discovered a universal dependence of r on n, namely an approximately linear dependence of $r^{-1}$ on $n^{-1}$, irrespective of the chain backbone structure or side groups. This discovery clarifies the previous results\cite{rAlQ3,rmeasurement1,rmeasurement2,nature,Friendnature,Bredas} and indicates that polymers with long CL are superior materials for application as active layers in $\pi$-conjugated OLEDs. This finding may also guide theoretical studies aiming to understand the underlying physics behind the spin-dependent exciton formation and its relation to electron correlation in $\pi$-conjugated materials.

To determine the ratio of the spin-dependent exciton formation cross-sections we have employed the PA and PADMR spectroscopies. The PA technique has been widely used in $\pi$-conjugated materials for studying long-lived photoexcitations having associated PA bands at subgap energies\cite{Wudl,Willi,Heeger,Handbook}. Two light beams are used in PA; one beam to excite the sample film over the optical gap and the other to probe the modulated changes, $\Delta T$ in the optical transmission, T. For excitation we used an Ar$^+$ laser beam that was modulated with a chopper. A combination of various incandescent lamps (xenon, tungsten-halogen and a glowbar), diffraction gratings, optical filters, and solid state detectors (silicon, germanium and indium-antimonide) was used to span the probe photon energy, $\hbar \omega$ between 0.3 and 3 eV. The PA spectrum, ($\Delta \alpha (\omega )$) was obtained by dividing $\Delta T/T$, where $\Delta T$ was measured by a phase sensitive technique referenced to the excitation modulation, and $\Delta \alpha = -d^{-1}\Delta T/T$ where d is the film thickness. 

The effect of spin-dependent recombination on the PA bands in the photomodulation spectrum was studied by the PADMR technique\cite{PADMR,PADMRa}. In this technique we measure the changes, $\delta T$ that are induced in $\Delta T$ by $\mu$-wave absorption in magnetic field, H in resonance with the Zeeman split spin-1/2 sublevels of photoinduced polarons in the film. $\delta T$ is proportional to $\delta N$ that is induced in the polaron density, N due to changes in the polaron pair recombination rates. Two types of PADMR spectra are possible: the H-PADMR spectrum in which $\delta T$ is measured at a fixed probe wavelength, $\lambda$ as the magnetic field H is scanned, and the $\lambda$-PADMR spectrum in which $\delta T$ is measured at a constant H in resonance, while $\lambda$ is scanned. We note that the PADMR setup allows both PA and PADMR spectra to be measured under identical conditions, so that the fractional change $\delta T/\Delta T$ is obtained with high precision and confidence.

For illustration of our method we choose a soluble oligothiophene, namely a dodecathiophene derivative (12T, or n = 12, for the molecular structure see Fig. 1(a), inset). In Fig. 1(a) we show the PA spectrum of an amorphous 12T film measured at 80K. The PA bands labeled $P_1$ and $P_2$ in the spectrum are spectral signatures of photogenerated long-lived polarons, both $P^+$ and $P^-$. The PA band marked $T_1$ is due to triplet exciton optical transitions\cite{triplet}. In pristine 12T film the band $T_1$ dominates the PA spectrum, whereas the polaron bands appear only as small shoulders. For the present measurements, where the polaron dynamics is essential we thus substantially enhanced the polaron bands in the PA spectrum by a photooxidation process, which is known to increase charge photogeneration in oligomer films\cite{Rothberg}. This was done by exposing the 12T film at ambient temperature to a UV-filtered beam from a 450-Watt xenon lamp for a three-hour period.

\begin{figure}[h]
\epsfig{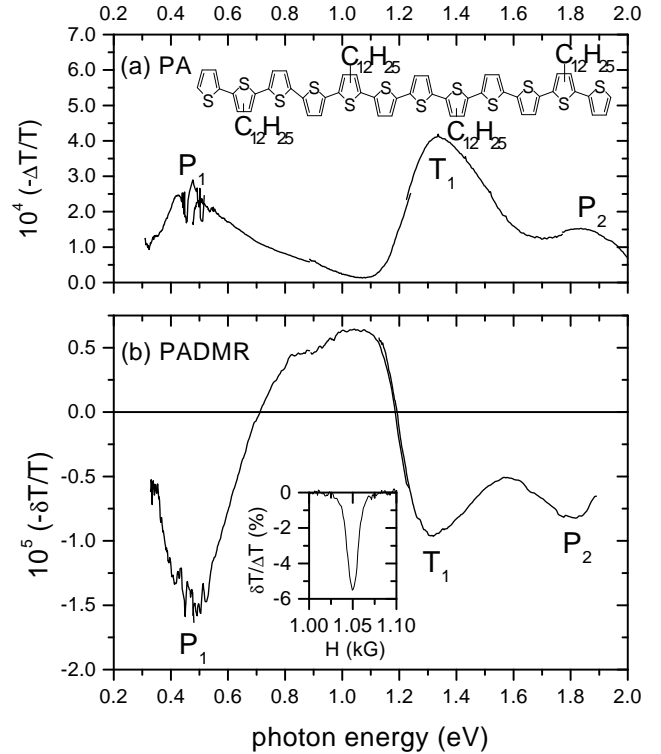}
\caption{(a) The photoinduced absorption (PA) spectrum of 12T, the molecular structure of which is shown in the inset; (b) the PA-detected magnetic resonance (PADMR) spectrum at magnetic field H=1.05 kG corresponding to S=1/2 resonance (see inset in (b)). Both spectra (a) and (b) show two bands ($P_1$ and $P_2$) due to polarons, and one band ($T_1$) due to triplet absorption. The PA was measured at 80 K, the excitation was at the 488 nm line (500 mW) of an $Ar^+$ laser; the PADMR spectrum was measured at 10 K.}
\end{figure}

In PA under steady state excitation conditions charge-transfer (CT) reactions occur between neighboring photogenerated $P^+$ and $P^-$. The CT reaction rate $R_P$ between spin parallel pairs (for the two spin up and down directions) is proportional to $2\sigma_T$, whereas the CT reaction rate $R_{AP}$ between spin antiparallel pairs (both spin up-down and down-up) is proportional to $(\sigma_S + \sigma_T$), where the proportionality constant is the same in both cases\cite{nature}. Since $\sigma_S > \sigma_T$ in $\pi$- conjugated compounds, then $R_{AP} > R_P$, and spin polarization of the recombining polaron pairs is built up over time, such that spin parallel pairs prevail at steady state conditions. Under saturated magnetic resonance conditions the Zeeman split sublevels become equally populated, so that the polaron pair densities with parallel and antiparallel spins are equal. Thus the PADMR measurements detect a reduction $\delta N$ (which is proportional to $\delta T$) in the polaron density N (which is proportional to $\Delta T$), since slowly recombining parallel pairs are converted to more efficiently recombining antiparallel pairs. At the same time the density of triplet excitons also decreases as a result of the decrease in the density of parallel polaron pairs. The quantity $\delta T/\Delta T$ is thus a direct measure of the fractional change in the overall polaron pair density, $\delta N/N$. For distant pair kinetics under saturation conditions we have\cite{Handbook} $\delta N/N = -(R_P-R_{AP})^2/(R_P+R_{AP})^2$, which when using the proportionality relations above between $R_P$, $R_{AP}$ and $\sigma_S$ and $\sigma_T$ translates to\cite{nature}:

\begin{equation}
r = \sigma_S/\sigma_T = \frac{1+3(\delta T/\Delta T)^{1/2}}{1-(\delta T/\Delta T)^{1/2}}
\end{equation} 
Thus the combination of PA and spin-1/2 PADMR spectroscopies gives a direct measure of r.
 
The spin 1/2 PADMR spectrum of 12T (Fig. 1(b)) clearly shows the negative magnetic resonance response at the PA bands $P_1$ and $P_2$, which are due to a reduction in the photogenerated polaron density. In addition we also see a negative response at the triplet band $T_1$, which confirms our assumption that $r > 1$. The resonant reduction of triplets should be compensated by an increase in the singlet exciton population and thus an increase in the photoluminescence under magnetic resonance condition was measured\cite{Shinar}. Also we tentatively assign the positive PA band in the $\lambda$-PADMR spectrum around 1 eV to a singlet exciton absorption band caused by excess singlet excitons that result from the excess spin anti-parallel polaron pairs created upon the $\mu$-wave absorption. However more work, especially in the ps time domain has to be completed to confirm this assignment.

$\delta T/\Delta T$ was determined for 12T to be 5.5\% at resonance (Fig. 1 (b) inset); using Eq. (1) this translates to r = 2.2.  Similar measurements were also completed for three other oligothiophenes, where r was also determined. The results are summarized in Fig. 2, where r for all four oligothiophenes and regio-regular poly(3-hexylthiophene) [RR-P3HT]  is plotted vs. 1/n, where n is the number of the five-member rings in the oligomer; we assumed 1/n = 0 for RR-P3HT due to the lamellar morphology of this film\cite{Sirringhaus,science}. From Fig. 2 it is seen that r decreases for short oligomers, which by extrapolation reaches $r \approx 1$ for the monomers. However we show below that our results are more general, where r appears to decrease with the CL in $\pi$-conjugated compounds largely independent of the polymer backbone structure.  

\begin{figure}[h]
\epsfig{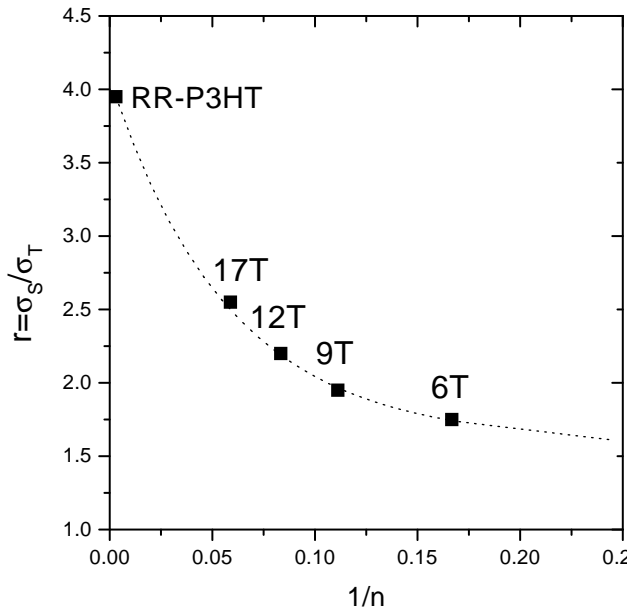}
\caption{The ratio r=$\sigma_S/\sigma_T$ of spin-dependent exciton formation cross sections for four oligothiophenes and regio-regular poly(3-hexylthiophene) [RR-P3HT] as a function of 1/n, where n is the number of the five-member rings in the oligomer; we assumed 1/n = 0 for RR-P3HT due to the lamellar morphology of this film. The line through the data points is a guide to the eye.}
\end{figure}

For this purpose we first demonstrate a direct and universal spectroscopic method that we found to obtain the CL of $\pi$-conjugated compounds. Obviously oligomers have a well-defined CL that is equal to the number of the repeat units in the chain; however polymers may have a very long chain-length. Nevertheless it is well established that the chemical conjugation and thus the electronic extent of the $\pi$-electron wave functions are only defect-free on a length-scale of the much smaller CL\cite{knupfer}. We may therefore deduce the polymer CL by inspecting the optical transitions of intrachain electronic excitations that are constrained by the CL, in much the same way as energy levels of a "particle in a box". Fig. 3 shows the dependence of the lower absorption band $P_1$ of polarons vs. 1/n for a variety of singly oxidized oligomers; the experimental values were taken from the literature\cite{P1data}. We emphasize that Fig. 3 contains data for many oligomers of different length as well as for three different $\pi$-conjugated systems, namely oligothiophenes, oligophenyls and oligophenylene-vinylenes. It is seen that the band $P_1$ depends linearly on 1/n irrespective of the $\pi$-conjugated backbone system. In fact the dependence of $P_1$ on 1/n is universal and therefore can be used to infer the average CL of many $\pi$-conjugated polymer films. We may therefore deduce the polymer CL in our studies from the $P_1$ peak energy in the PA spectrum. This can be readily done since the magnitude and photon energy of the $P_1$ band was carefully measured in our PA measurements.

\begin{figure}[h]
\epsfig{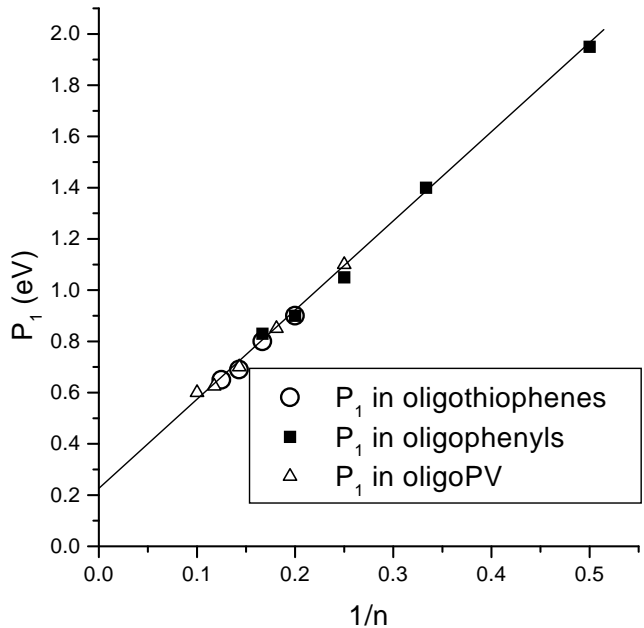}
\caption{The peak photon energy of low energy polaron transition, $P_1$ in singly oxidized oligomers as a function of the inverse conjugation length (CL). The number n denotes a CL that corresponds to an oligothiophene with n rings. The data for the different oligomer classes were taken from [20].The line through the data points is a linear fit.}
\end{figure}

In order to show the universality of the function r(n) in $\pi$-conjugated materials we replotted in Fig. 4 our results of r vs. the optical gap\cite{nature}, in the form of $r^{-1}$ vs. $P_1$; for completeness we have also included in this plot our present results in oligothiophenes. Since $P_1$ is a linear function of 1/n we may actually plot $r^{-1}$ vs 1/n as is also shown in Fig. 4 (upper axis). Amazingly we discover a universal behavior of r(n), namely that $r^{-1}$ depends linearly on 1/n irrespective of the chain backbone structure. We note that, as is apparent from Fig. 4, we encounter negative values for 1/n for several polymers that have the lowest $P_1$ transitions (see \cite{negative} for a discussion).

\begin{figure}[h]
\epsfig{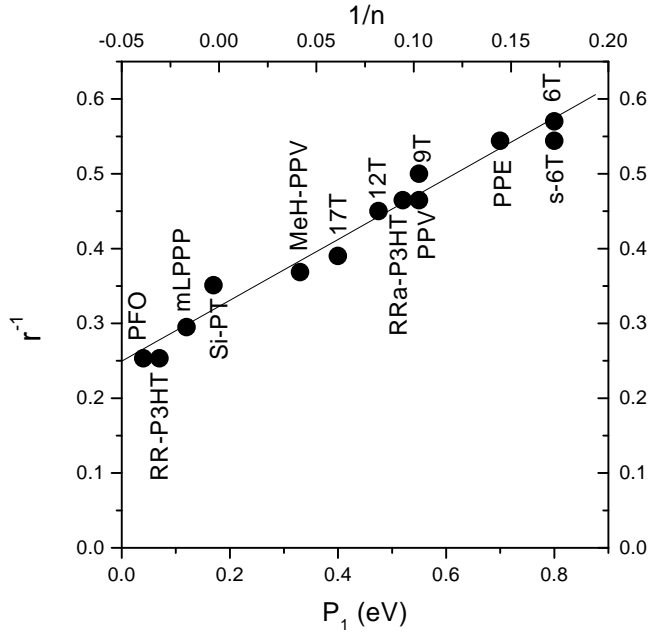}
\caption{The ratio $r^{-1}$=$\sigma_T/\sigma_S$ of spin-dependent exciton formation cross sections in various polymers and oligomers as a function of the peak photon energy of the $P_1$ transition (lower x-axis, see text). $r^{-1}$ is also shown as a function of the inverse conjugation length 1/n (upper x-axis), which was determined from $P_1$ by the linear extrapolation of the data shown in Fig. 3 (see text). The line through the data points is a linear fit.}
\end{figure}

By linearly extrapolating the universal dependence of $r^{-1}$ on $n^{-1}$ we get $r \approx 1$ for $n \approx 2 (\approx 7 \AA$\cite{OT}). We conclude therefore that exciton formation should become spin-independent for oligomers of length $\approx 7 \AA$ or less. This shows that $\sigma_S \approx \sigma_T$ for small molecules, in agreement with the previous results in AlQ$_3$ \cite{rAlQ3} and a Platinum containing phenylene-ethylene-type monomer \cite{Friendnature}, which were deduced more directly from the EL quantum yield of OLEDs. Since the maximum quantum yield of OLED is related to r, then our results show that if all other factors are equal the quantum yield of polymer OLEDs is larger than that of OLEDs based on small molecules; this constitutes a clear advantage for the polymers.

We advance several speculations to explain the very clear r(n) dependence. Firstly we note that the wave-function extent of singlet excitons was measured to be as large as the chain CL\cite{knupfer}. On the contrary the wave-function extent of triplet excitons is always very confined: Indeed an average separation of electron and hole in the triplet state of PPV was measured to be $\approx 3.2 \AA$ \cite{RonPRB}. Therefore the difference between the singlet and triplet excitons significantly decreases with 1/n\cite{exciton}, and this may explain the reason that r decreases with 1/n. Alternatively the variation of r with n may be due to the variation of the Langevin capture mechanism of charge recombination with CL\cite{Friendnature}. When the Langevin capture radius, $R_L >> na$, where a is the monomer length, then the recombining polarons reside on separate molecules and therefore their wave-function overlap is small. This, in turn reduces the exchange energy that figures in the triplet energy, thus reducing the difference between polaron pair recombination with parallel or antiparallel spins. However the apparent linear dependence of $r^{-1}$ on 1/n remains a mystery.

In summary we have measured the formation cross-section ratio, r of singlet and triplet excitons that are generated in $\pi$-conjugated materials from polaron pairs, using the PA and PADMR spectroscopies. We deduced r from the change in the spin dependent recombination of polaron pairs in a wide variety of $\pi$-conjugated oligomer and polymer films. In oligothiophenes we found that r increases with the chain length. Based on a universal relationship between the low-energy polaron absorption and the chain length in oligomers, we discover that $r^{-1}$ is linearly proportional to the inverse CL in both polymers and oligomers. This may be due to the suppression of the distinctively contrasting properties of triplet and singlet excitons at small CL. 

We thank Dr. Mazumdar for useful discussion and encouragement to complete this work. We also thank Drs. Scherf for kindly supplying the mLPPP polymer, Bradley for the PFO polymer and Barton for supplying the Si-bridged PT and PPE polymers; we also thank Drs. DeLong and Chinn for preparing the DOO-PPV and PPV polymers. The work at the University of Utah was partially supported by DOE ER-45490.


\begin{references}
\bibitem{negative} In the present scheme these polymers would therefore be assigned a conjugation-length "larger than infinity". This situation was actually also encountered previously by several authors \cite{science,TwoD} in RR-P3HT, which was shown to form two-dimensional lamellae \cite{Sirringhaus,science}. It was indeed shown\cite{science} that exceptionally low photon energy for the $P_1$ band results from a delocalization of polarons not only along the conjugated polymer backbone, but also in the two-dimensional lamellae. This suggests an interpretation of the "larger than infinite" CL as a situation where excitations extend not only along the backbone, but also along the perpendicular directions. It is however not entirely clear whether the linear extrapolation of the data presented in Fig. 3 to 1/n=0 is exactly valid. In any case polymers with low $P_1$ band correspond to highly conjugated systems.
\end{references}
\end{document}